\documentclass[twocolumn,floatfix,prl,showpacs,superscriptaddress]{revtex4}

\usepackage{graphicx}
\usepackage{amssymb}
\usepackage{amsmath}
\usepackage{color}
\usepackage{psfrag}
\usepackage{epsfig}
\usepackage{bbm}
\usepackage{bm}
\usepackage{simplewick}
\usepackage{dsfont}

\newcommand{\ket}[1]{| #1 \rangle}
\newcommand{\bra}[1]{\langle #1 |}
\newcommand{\rb}[1]{\left( #1 \right)}

\newcommand{\ew}[1]{\langle #1 \rangle}
\newcommand{\beq}{\begin{eqnarray}}
\newcommand{\eeq}{\end{eqnarray}}

\newcommand{\op}[2]{| #1 \rangle \langle #2 |}

\newcommand{\eq}[1]{Eq.~(\ref{#1})}
\newcommand{\kett}[1]{| #1 \rangle\!\rangle }

\newcommand{\opp}[2]{| #1 \rangle\! \rangle\langle\! \langle #2 |}

\begin{document}

\title{Non-Markovian effects in the quantum noise of interacting nanostructures}

\author{D. Marcos}
\affiliation{Theory and Simulation of Materials, Instituto de Ciencia de Materiales de Madrid (ICMM-CSIC), Cantoblanco 28049, Madrid, Spain}
\affiliation{Institute for Quantum Optics and Quantum Information of the Austrian Academy of Sciences, A-6020 Innsbruck, Austria}
\affiliation{Institute for Theoretical Physics, University of Innsbruck, Technikerstr. 25, A-6020 Innsbruck, Austria}

\author{C. Emary}
\affiliation{Institut f\"ur Theoretische Physik, Hardenbergstr. 36, TU Berlin, D-10623 Berlin, Germany}

\author{T. Brandes}
\affiliation{Institut f\"ur Theoretische Physik, Hardenbergstr. 36, TU Berlin, D-10623 Berlin, Germany}

\author{R. Aguado}
\affiliation{Theory and Simulation of Materials, Instituto de Ciencia de Materiales de Madrid (ICMM-CSIC), Cantoblanco 28049, Madrid, Spain}

\begin{abstract}
We present a theory of finite-frequency noise in non-equilibrium conductors.
It is shown that Non-Markovian correlations are essential to describe the physics of quantum noise. In particular, we show the importance of a correct treatment of the initial system-bath correlations, and how these can be calculated using the formalism of quantum master equations. Our method is particularly important in interacting systems, and when the measured frequencies are larger that the temperature and applied voltage. In this regime, quantum-noise steps are expected in the power spectrum due to vacuum fluctuations. This is illustrated in the current noise spectrum of single resonant level model and of a double quantum dot --charge qubit-- attached to electronic reservoirs. Furthermore, the method allows for the calculation of the single-time counting statistics in quantum dots, measured in recent experiments.
\end{abstract}
 \maketitle

\section{Introduction}

Vacuum fluctuations are one of the most intriguing consequences of the quantum theory. In electronic systems, they manifest as electron-hole creation/annihilation processes in a time given by the Heisenberg uncertainty relation, $t\sim1/\omega$, being $\omega$ the measuring frequency. 
In order for these processes to be seen, other types of fluctuations must be overcome. For example, a system in thermodynamic equilibrium must be at a temperature $T$ much smaller than this frequency, and in a system driven out of equilibrium, such as a mesoscopic conductor subject to an applied voltage $V$, the quantum-noise regime (QNR) reads $\hbar\omega\gg k_B T, eV$.
Zero-point fluctuations in quantum-transport systems were first measured by Schoelkopf and collaborators \cite{Schoelkopf97} through the current-noise spectrum \cite{steady}
\beq
S^{(2)}(\omega)= \int_{-\infty }^{\infty }d\tau e^{-i\omega\tau}\langle\{\hat{I}(t+\tau),\hat{I}(t)\}\rangle_c,
\eeq
which reveals valuable information beyond that contained in the dc current \cite{bla00,gala03nag04pil04sal06,Aguado-Brandes}. Among the various methods to calculate $S^{(2)}(\omega)$, quantum master equations (QMEs) are particularly attractive because of their simplicity and generality for treating dissipative dynamics of interacting systems \cite{Choi01,Gurvitz,Ruskov,Aguado-Brandes,Emaryetal07}. Typically, the Markovian approximation (MA) in the system-reservoir coupling is employed. This, however, fails in describing the noise spectrum in the QNR \cite{Marcosetal10}, and although there have been a few attempts to go beyond the MA in the context of QMEs \cite{Engel-Loss}, a complete noise theory is yet lacking.

\begin{figure}[t]
\center
\includegraphics[width=3.5in]{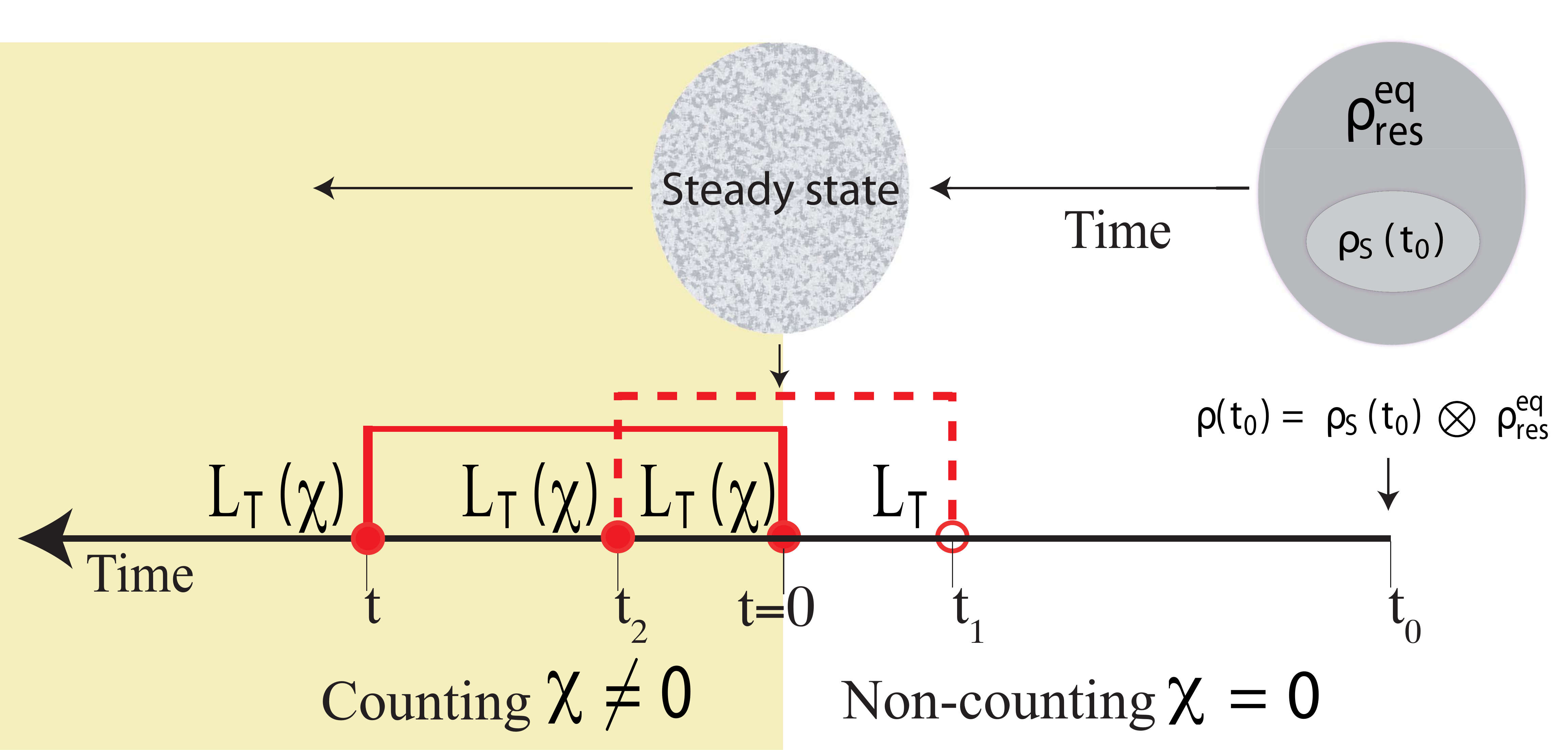}
\caption[Non-Markovian time evolution]{\label{Fig1paperNM} 
(color online). Schematics of counting: The density operator evolves from the initial separable state at time $t_0$ (represented by two distinct ellipses) until it reaches a steady state at time $t=0$, where it is no longer in a product state (single ellipse). At time $t=0$ counting begins. The shading highlights the time interval where counting is effective. Full circles denote tunnel vertices with counting factors $\chi\neq 0$, empty circles denote standard tunneling vertices ($\chi=0$). Contractions between tunneling events in counting and non-counting intervals (dashed over-line) give rise to $\Gamma(\chi,z)$, while contractions within the counting interval (solid over-line) give rise to the self-energy $\Sigma(\chi,z)$.}
\end{figure}

In this paper we present such a theory. Our method allows the calculation of the current and voltage noise spectrum of a system described by a generic non-Markovian QME, and can be applied to the increasing number of experiments exploring the QNR \cite{highfreqexp}. The theory naturally contains the physics of vacuum fluctuations, for which a proper inclusion of initial system-bath correlations is essential. Furthermore, the method enables to determine the charge-noise spectrum 
\beq \label{chargenoisespectrum}
S^{(2)}_Q(\omega)= \int_{-\infty }^{\infty }d\tau
e^{-i\omega\tau}\langle\{Q(\tau),Q(0)\}\rangle_c,
\eeq 
as it is shown for a single resonant level (SRL) model. This noise dictates the back-action when the conductor is used as a detector of another quantum system \cite{backaction1}. The technique is used to study the full noise spectrum of a double quantum dot charge qubit in the hitherto unexplored QNR. As we will see, in this regime transport fluctuations are mediated by the zero-point dynamics, showing a series of steps at frequencies corresponding to resonant processes in the system.

\section{Theory}

Here we consider phenomena that can be described by the general QME 
\beq \label{LiouvilleEq}
\dot{\rho}(t)={\cal L}\rho(t), 
\eeq
where ${\cal L}$ is the Liouvillian, that governs the evolution of the density operator (DO), $\rho$, describing the dynamics of the total system. Specifically, we focus on the case in which a central system exchanges particles with a bath, and this exchange is amenable to the counting of particles. We will take here the case of transport through a central quantum coherent system, attached to fermionic contacts. The Hamiltonian of the system is of the form
${\cal H} = {\cal H}_\mathrm{S} + {\cal H}_\mathrm{R}
+{\cal H}_\mathrm{V}$. Here ${\cal H}_\mathrm{S} =\sum_{a} E_{a} \op{a}{a}$ is the central-system Hamiltonian, with $E_{a}$ the energy of the $N_a$-electron \emph{many-body} eigenstate $\ket{a}$.
The left/right reservoirs (at equilibrium with chemical potentials $\mu_{L/R}=E_F\pm eV/2$) are described by 
${\cal H}_\mathrm{R} = \sum_{k,\alpha\in L,R} \varepsilon_{k\alpha}
c^\dag_{k\alpha} c_{k\alpha}$, with $\varepsilon_{k\alpha}$ the
energy of the $k$-th mode in lead $\alpha$. The tunnelling
Hamiltonian is given by ${\cal H}_V=
\sum_{k\alpha m}V_{k\alpha m} c^\dag_{k\alpha} d_m
  + \mathrm{H.c.}=\sum_{k\alpha m} \sum_{a,a'}V_{k\alpha m} c^\dag_{k\alpha} \langle a|d_m|a'\rangle\op{a}{a'}
  + \mathrm{H.c.}$, where $c^\dag_{k\alpha}$ creates an electron with momentum $k$ in reservoir $\alpha$ and $d_m$ is the annihilation
operator for the single-particle level $m$ in the central-system. $V_{k\alpha m}$ is a tunnelling amplitude and $e=\hbar = k_B=1$ throughout the text. 
Under the previous Hamiltonian, the DO evolves according to equation (\ref{LiouvilleEq}), with ${\cal L}\bullet \equiv -i \left[{\cal H}_\mathrm{S}+{\cal H}_\mathrm{R}+{\cal H}_\mathrm{V},\bullet\right]\equiv({\cal L}_\mathrm{S} + {\cal L}_\mathrm{R} + {\cal L}_\mathrm{V})\bullet$. 
We are interested in the central-system dynamics, for which we consider the reduced system DO $\rho_\mathrm{S}(t)\equiv\mathrm{Tr}_\mathrm{R}\{\rho(t)\}$. 
If we choose $t_0$ to be the time at which system and reservoirs are in a separable state,
$\rho(t_0) =
\rho_\mathrm{S}(t_0) \otimes \rho_\mathrm{R}^\mathrm{eq}$, with
$\rho_\mathrm{S}(t_0)$ arbitrary and $
\rho_\mathrm{R}^\mathrm{eq}$ the equilibrium bath state, the evolution of $\rho_\mathrm{S}(t)$ in the Laplace space is given by
\beq \label{rhoz}
  {\rho}_\mathrm{S}(z)
  =\mathrm{Tr}_\mathrm{R}
  \left\{
    \left[z-{\cal L}\right]^{-1}
    \rho_\mathrm{S}(t_0) \otimes \rho_\mathrm{R}^\mathrm{eq}
  \right\}
  =\Omega_0(z){\rho}_\mathrm{S}(t_0).
\eeq 
Here, we find the propagator $\Omega_0(z) \equiv \left[z-{\cal W}(z)\right]^{-1}$, with kernel ${\cal W}(z) = {\cal L}_\mathrm{S} +
\Sigma(z)$, being $\Sigma(z)$ the non-Markovian (NM) self-energy, and whose form can be derived using the expansion
\beq
\frac{1}{z-{\cal L}}= \frac{1}{z-{\cal L}_\mathrm{S} -{\cal L}_\mathrm{R}}\sum_{k=0}^{\infty} \left( {\cal L}_\mathrm{V}\frac{1}{z-{\cal L}_\mathrm{S} -
{\cal L}_\mathrm{R}}\right)^k.
\eeq
This gives 
\beq
\Sigma(z)=\mathrm{Tr}_\mathrm{R}
  \left\{ {\cal L}_\mathrm{V}\frac{1}{z-{\cal L}_\mathrm{S} -
{\cal L}_\mathrm{R}}{\cal L}_\mathrm{V}\rho_\mathrm{R}^\mathrm{eq}\right\} +
  \ldots
\eeq 
Technical details on how to evaluate this expression \cite{Schoeller09,CE09cot,CE10SCLPT1} are not relevant for the main discussions and are given in appendix \ref{appKernel}.

\subsection{Cumulant generating function}

Our goal here is, given Eq.~(\ref{rhoz}), to derive a formula for the cumulant generating function (CGF) in terms of known quantities such as the self-energy. This will allow us to calculate NM current correlations up to arbitrary order at zero frequency. Furthermore, we aim to give an expression for the NM finite-frequency noise correlation function. If the transfer of electrons between system and reservoirs is amenable to counting, the full counting statistics of the number of transferred electrons $n$ can be studied with the DO formalism. To do this, we unravel $\rho_\mathrm{S}(t)$ in terms of this continuous projective measurement: $\rho_\mathrm{S}(t)=\sum_n \rho_\mathrm{S}^{(n)}(t)$, similarly to how this is done in quantum optics \cite{Cook}. The probability distribution of having $n$ transfers after time $t$ is given by $P(n,t)=\mathrm{Tr}_\mathrm{S}\{\rho_\mathrm{S}^{(n)}(t)\}$, and the corresponding CGF is ${\cal F}(\chi;t)\equiv\ln\sum_{n=-\infty}^{\infty}P(n,t) e^{in\chi}$.
This allows to calculate the {\it k}-th order cumulant of the current distribution as $\langle I^k(t)\rangle_c=\langle \dot{n}^k(t)\rangle_c=\frac{d}{dt}\frac{\partial^k {\cal
F}(\chi,t)}{\partial(i\chi)^k}|_{\chi\rightarrow
0}$. In practice, counting in lead $\alpha$ can be effected by adding $\chi_{\alpha}$ to the tunneling Liouvillian ${\cal
L}_\mathrm{T}$ through the replacement $V_{k \alpha m}\to V_{k
\alpha m} e^{i p \chi_{\alpha}/2}$ \cite{levitov04},
where $p=+/-$ is the Keldysh index corresponding to the forward/backward time branch.
Derivatives with respect to different counting fields, e.g. $\chi_L$, $\chi_\mathrm{R}$, allow us to obtain also cross correlations of currents flowing through different contacts. In the following, the lead-dependence of the counting field will be considered implicit. Let us try to relate this CGF (or alternatively the moment generating function ${\cal G}\equiv e^{\cal F}$) with a general NM evolution. In the $\chi$-space, the density operator
$ \rho_\mathrm{S}(\chi,z) \equiv\sum_{n=-\infty}^{\infty} \int_0^{\infty} dt \rho_\mathrm{S}^{(n)}(t)e^{i\chi n-zt}$ follows the evolution $\rho_\mathrm{S}(\chi,z)
  =\Omega(\chi,z)\rho_\mathrm{S}(0)
$, with $\Omega(\chi,z)\equiv [z-{\cal W}(\chi,z)]^{-1}$, and ${\cal W}(\chi,z)= {\cal L}_\mathrm{S} +
\Sigma(\chi,z)$. To lowest order we have
\beq \label{1pointSE}
\Sigma(\chi,z)=\mathrm{Tr}_\mathrm{R}
  \left\{ {\cal L}_\mathrm{V}(\chi)\frac{1}{z-{\cal L}_\mathrm{S} -
{\cal L}_\mathrm{R}}{\cal L}_\mathrm{V}(\chi)\rho_\mathrm{R}^\mathrm{eq}\right\}.
\eeq
For later use, we also introduce the two-point self-energy 
\beq \label{2pointSE}
\Pi(\chi_2,\chi_1,z)=\mathrm{Tr}_\mathrm{R}
  \left\{ {\cal L}_\mathrm{V}(\chi_2)\frac{1}{z-{\cal L}_\mathrm{S} -
{\cal L}_\mathrm{R}}{\cal L}_\mathrm{V}(\chi_1)\rho_\mathrm{R}^\mathrm{eq}\right\}.
\eeq
Obviously, we have $\Pi(\chi,\chi,z)=\Sigma(\chi,z)$, and $\Sigma(\chi=0,z)= \Sigma(z)$. Explicit expressions for Eqs. (\ref{1pointSE}) and (\ref{2pointSE}) are given in appendix \ref{appKernel}.

In the widely used Born-Markov approximation, the state at which counting begins (say $t=0$) can be taken to be $\rho_\mathrm{S}(0)\otimes \rho_\mathrm{R}^\mathrm{eq}$. However, to consider NM corrections, the state at time $t=0$ can no longer be considered as a separable state, as it contains initial system-bath correlations.
To account for these, we explicitly divide the time evolution into two intervals (see Fig.~\ref{Fig1paperNM}). The evolution from $t_0$ (time at which system and reservoirs are separable) to $t=0$ is given by $\frac{1}{z_0-{\cal L}}$, while the evolution from $t=0$ to $t$ is given by $\frac{1}{z-{\cal L}(\chi)}$. Doing this we obtain the moment generating function (MGF):
\beq
  \label{inhomo1} {\cal G}(\chi;z)
  =
  z_0\mathrm{Tr} \left\{
    \frac{1}{z-{\cal L}(\chi)}\frac{1}{z_0-{\cal L}}
    \rho_\mathrm{S}(t_0) \otimes \rho_\mathrm{R}^\mathrm{eq}
  \right\}.
\eeq
Here $z$ is the conjugate frequency to $t$, and $z_0$ to $-t_0$. We will take $t_0\to -\infty$, which implies $z_0\to 0^-$ (henceforth implicit). The trace in (\ref{inhomo1}) refers to the full trace (system plus bath degrees of freedom).
Using geometric expansions of $\frac{1}{z-{\cal L}(\chi)}$ and $\frac{1}{z_0-{\cal L}}$, and performing the trace over the reservoirs, we get 
\beq\label{inhomo2}
  {\cal G}(\chi;z)
  =
   \Big\langle
\frac{1}{z-{\cal L}_\mathrm{S}-\Sigma(\chi,z)}\left(\mathds{1}+ \Gamma(\chi,z)\right)\Big\rangle.
\eeq
In this equation, $\ew{\ldots} \equiv \mathrm{Tr}_{\mathrm{S}}\left\{\ldots \rho_\mathrm{S}^\mathrm{stat}\right\}$, where we have taken $\rho_{\mathrm{S}}(0)=\rho_{\mathrm{S}}^{\mathrm{stat}}$, as we are interested in fluctuations around the stationary state. This can be obtained either as $\rho_\mathrm{S}^\mathrm{stat}=\lim_{z\to 0} z \rho_\mathrm{S}(z)$ in equation (\ref{rhoz}), or solving ${\cal W}(0)\rho_\mathrm{S}^\mathrm{stat}=0$. The inhomogeneous term $\Gamma(\chi,z)$ in Eq.~(\ref{inhomo2}) is given by
\beq
  \Gamma(\chi;z) =\frac{1}{z}\{\Pi(\chi,0,z_0)-\Pi(\chi,0,z)\}
  +\ldots
  \label{Gamma}
  \eeq
Eqs. (\ref{inhomo2}) and (\ref{Gamma}) are the first main formal result of the paper. As we shall show below, the inclusion of $\Gamma(\chi,z)$ in the MGF is crucial to account for NM physics and quantum noise. Importantly, $\Gamma(\chi,z)$ cannot, in general, be cast in the form of a self-energy, since only \emph{one} of the two vertices (i.e. tunneling Liouvillians) contains a counting field $\chi$. 
Notice that Eq.~(\ref{Gamma}) extends the particular form of the inhomogeneity $\Gamma(\chi; z)= \frac{1}{z}\{\Sigma(0,0)-\Sigma(0,z)\}$, which appears in \cite{Flindt08}. This is only valid 
for a system with NM dynamics but with Markovian coupling with the bath in which counting is performed, and as a result, quantum fluctuations due to the Fermi contacts are not captured in this case.

\subsection{Noise spectrum}

From the MGF (\ref{inhomo2}), together with (\ref{Gamma}), we can derive a general equation for the noise spectrum. To this end we make use of the MacDonald's formula \cite{Flindt08,Lambert}
\beq
S^{(2)}(\omega) &=& \omega\int_0^{\infty} dt \mathrm{sin}(\omega t)\langle I^2(t)\rangle_c \nonumber\\ &=& -\frac{\omega^2}{2}
\frac{\partial^2}{\partial{(i\chi)^2}}\left[\mathcal{G}(\chi,z=i\omega)
+(\omega\rightarrow-\omega)\right]\Big|_{\chi\rightarrow 0}, \nonumber\\
\eeq
and obtain
\begin{widetext}
\beq
S^{(2)}(\omega) = \left[
   \ew{
     {\cal J}^{II}(i\omega,i\omega_0)
   }
   +
   \ew{
     {\cal J}^{I}(i\omega,i\omega) \Omega_0(i\omega){\cal J}^{I}(i\omega,i\omega_0)
   } \right]
   +(\omega \leftrightarrow -\omega)
  \label{Sz}
  ,
\eeq 
with $\omega_0\to 0$ and
\beq
{\cal J}^{II}(z,z_0) &\equiv&
    \frac{\partial^2}{\partial (i\chi_2)\partial (i\chi_1)}
    \Pi(\chi_2,\chi_1,z)
  \Big|_{\chi_2,\chi_1\to 0}
  +
  \ldots
  \label{J2nonMarkovian}
\\
  {\cal J}^{I}(z,z') &\equiv&  
  \frac{\partial}{\partial (i\chi)}
  \left[
    \Pi(0,\chi,z)+\Pi(\chi,0,z')
  \right]\Big|_{\chi\to 0}
  +
  \ldots
    \label{J1nonMarkovian}
\eeq
\end{widetext}
Eq. (\ref{Sz}), together with (\ref{J2nonMarkovian}) and (\ref{J1nonMarkovian}), is the second main formal result of the paper. 
It is exact and agrees with previous approaches in the literature in the appropriate limits \cite{Engel-Loss,Braun06}. In particular, the Markovian result\cite{Marcosetal10} is recovered by neglecting the frequency dependence of the jump super-operators: $ {\cal J}^{II}(z,z_0)\rightarrow {\cal J}^{II}(0,0)$, $ {\cal J}^{I}(z,z')\rightarrow {\cal J}^{I}(0,0)$. 
The correct NM zero-frequency limit \cite{Flindt08} is also recovered. It is interesting to notice that Eq.~(\ref{inhomo2}) not only allows us to obtain the NM noise spectrum, but also single-time NM correlations to arbitrary order, $\ew{I^N(t)}_c$, $\ew{n^N(t)}_c$, $\ew{Q^N(t)}_c$, by simply taking derivatives with respect to the counting field.

We notice that the above derivation has focused on particle currents flowing through the barriers separating central system and leads. At finite frequencies this particle current is not conserved due to charge accumulations in the system, and the total current (particle plus displacement) needs to be considered to obtain the noise spectrum.
However, our results are general, and current conservation can be considered by the inclusion of the proper counting fields \cite{Marcosetal10} in $\Sigma(\chi,z)$ and $\Pi(\chi_2,\chi_1,z)$. Thus, particle, total, and charge noise (equivalently voltage noise for a capacitive system), can be calculated from Eq.~(\ref{Sz}). To this end, it is enough to consider respectively \cite{alphabeta} $\chi_L/\chi_R$, $\chi_{\mathrm{tot}}\equiv\chi_L+\chi_R$, and $\chi_{\mathrm{accum}}\equiv \beta\chi_L-\alpha\chi_R$, giving rise to different jump super-operators \cite{Marcosetal10}. 

\begin{figure*}
\center
\includegraphics[width=1.0\textwidth]{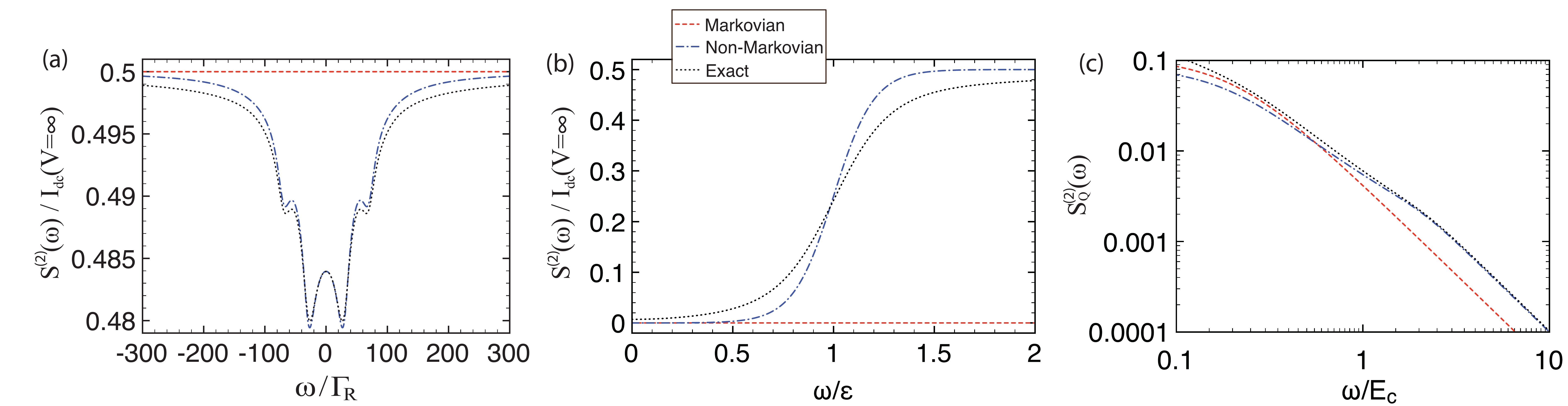}
\caption[Quantum noise spectra of a single resonant level model]{(color online). Quantum noise spectra of the SRL model ($\Gamma_L=\Gamma_R=1$ in all figures). a)~$S^{(2)}(\omega)$ as a function of frequency $\omega$ in the shot noise regime ($\varepsilon=20$, $V=100$, $T=4$). In this limit, the noise develops dips at $\omega=\pm|\varepsilon\pm\frac{eV}{2}|$ . b)~$S^{(2)}(\omega)$ as a function of frequency $\omega$ in the quantum noise regime ($\varepsilon=10$, $V\rightarrow 0$, $T=1$). In this limit, $S^{(2)}(\omega)$  develops a quantum noise step at $\omega=\varepsilon$. c)~Charge noise $S^{(2)}_Q(\omega)$ as a function of frequency $\omega$ of a single electron transistor acting as a detector ($E_C=V=10$).  When $\omega>E_C$, $S^{(2)}_Q(\omega)$ contains extra quantum noise contributing to backaction.}
\label{Fig1paperNMsup}
\end{figure*}

\section{Results}

\subsection{Single resonant level model}

We now use the formalism presented in the previous section to calculate the NM noise spectrum of a single resonant level model (equivalently of a single electron transistor with $E_C\gg k_BT$, being $E_C$ the charging energy, and with only two relevant charge states). The noise and charge spectrum of this system have already been calculated with a variety of techniques \cite{bla00,Engel-Loss,Johansson02}, and the exact solution is also well known \cite{Averin}. We therefore use this as a benchmark of our method. In the following we show the good agreement between our theory and the exact solution. In the QNR, these two, in contrast to the Markovian result, show quantum-noise steps due to vacuum fluctuations, as we will see. The Markovian and non-Markovian results we present here correspond to first order in perturbation theory (sequential tunneling) and in the following $S^{(2)}(\omega)$ refers to the `total' noise.

The SRL model is described by the Hamiltonian
\beq 
{\cal H}&=&\varepsilon |1\rangle\langle 1|+\sum_{k,\alpha\in L,R} \varepsilon_{k\alpha}
c^\dag_{k\alpha} c_{k\alpha}
\nonumber\\
&+&\sum_{k,\alpha\in L,R} V_{k\alpha} c^\dag_{k\alpha} |0\rangle\langle 1| + \mathrm{H.c.}
\label{srl}
\eeq
Here, each of the terms corresponds to central system, reservoirs, and tunneling respectively. 
The state $\ket{1}$ (occupied level), together with $\ket{0}$ (empty level) form the Hilbert space of the central system ($|0\rangle\langle 0|+|1\rangle\langle 1|=\mathbbm{1}$). This model, despite its simplicity, contains a great deal of interesting physics: In the context of mesoscopic systems, this Hamiltonian captures the physics of a quantum dot in which only one single level participates in transport (strong Coulomb Blockade regime). Also, it can be shown that there is an exact mapping between the SRL model and the spin-boson model (namely a quantum two-level system coupled with strength $\alpha$ to an Ohmic dissipative bosonic bath) at  $\alpha=1/2$. This mapping is actually an special case of the more general relation between the spin-boson model and the anisotropic Kondo model, for which  $\alpha=1/2$ is the exactly solvable point, the so called Toulouse limit of the Kondo problem \cite{mappingSRL}.

Fig.~\ref{Fig1paperNMsup}a shows the shot noise spectrum $S^{(2)}(\omega)$ of the total current through the system obtained with the non-Markovian formalism discussed in the previous section (blue dashed-dotted curve). We also plot the exact result \cite{Averin} (black dotted curve) and the one obtained after a Markovian approximation \cite{Marcosetal10} (red dashed curve). The agreement between the exact solution and the NM calculation is extremely good. Both develop dips at frequencies 
$\omega=\pm|\varepsilon\pm\frac{eV}{2}|$, and show a strong frequency dependence. As expected, and due to the mapping aforementioned, the shot noise spectrum in Fig.~\ref{Fig1paperNMsup}a agrees well with the one of a non-equilibrium Kondo model in the Toulouse limit \cite{Schiller-Hershfield}. In stark contrast, the Markovian solution is markedly different: it is frequency-independent and equals $S^{(2)}(\omega\rightarrow\infty)=\frac{\Gamma_L\Gamma_R}{2(\Gamma_L+\Gamma_R)}=\frac{\langle I\rangle}{2}$. Even at $\omega=0$, the MA deviates from the NM and exact solutions, which here fall practically on top of each other. In Fig.~\ref{Fig1paperNMsup}b, we explore the linear-response regime when the level is outside the bias voltage window. In this situation shot-noise is negligible, and quantum fluctuations are dominant in the spectrum for $\hbar\omega\gg k_BT$. The quantum noise step expected at $\omega=\varepsilon$ is fully captured by our NM approach, while here it becomes clear that the MA does not capture quantum noise physics. 

\begin{figure}[t]
\center
\includegraphics[width=2.5in]{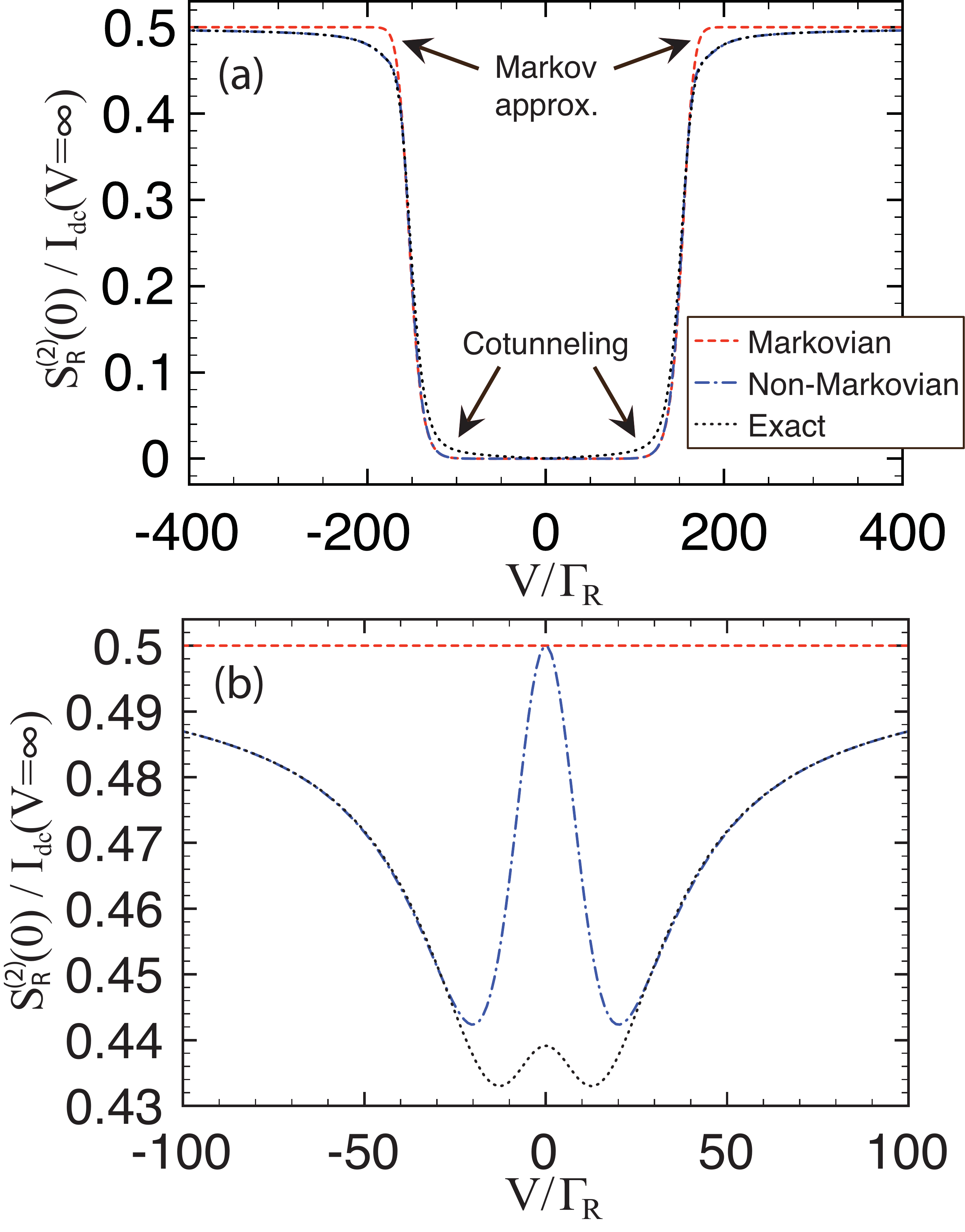}
\caption[Non-Markovian zero-frequency limit of the single resonant level model]{(color online). Zero-frequency limit of the non-Markovian theory. a)~Particle-current noise for $\Gamma_L=\Gamma_R$, $\varepsilon/\Gamma_R=80$, $T/\Gamma_R=4$ The noise suddenly increases when the level enters the voltage bias window. b)~$S^{(2)}_R(0)$ as a function of bias voltage $V$ for $\Gamma_L=\Gamma_R$, $\varepsilon=0$, $T/\Gamma_R=4$. While the Markovian approximation is flat at all voltages, the NM and exact solution show a structure that strongly differs in both for low voltages due to cotunneling processes.}
\label{Fig2paperNMsup}
\end{figure}

The richness of the SRL model can be further explored by noting that it also describes the physics of a single electron transistor (SET) with charging energy $E_C\gg k_BT$, and voltage such that only two charge states $|N\rangle$ and $|N+1\rangle$ are relevant. One can describe a SET in this regime with Eq. (\ref{srl}) by just making the substitutions \cite{SchoelkopfCH} $\varepsilon\rightarrow E_C$, $|0\rangle\rightarrow|N\rangle$ and $|1\rangle\rightarrow|N+1\rangle$. Let us derive the charge-noise spectrum (\ref{chargenoisespectrum}) of the SET. This problem has already been studied by Johansson {\it et al.} using a different formalism \cite{Johansson02}. As discussed in the previous section, $S^{(2)}_Q(\omega)$ can be found by considering the jump operators arising form the counting field $\chi_{\mathrm{accum}}\equiv\beta\chi_L-\alpha\chi_R$, being $\alpha$ and $\beta$ coefficients determining how the total current is partitioned between both left and right contacts \cite{Marcosetal10}. Alternatively, we can apply charge conservation: $I_L(t)-I_R(t)=\dot{Q}(t)$, being $I_{L/R}$ the current through the left/right lead and $Q$ the charge inside the well. This, together with the Ramo-Shockley partitioning theorem $I(t)=\alpha I_L(t) +\beta  I_R(t)$ to obtain
\beq
S^{(2)}_Q(\omega)=\frac{1}{\omega^2}\left[ S^{(2)}_L(\omega)+S^{(2)}_R(\omega)-S^{(2)}_{LR}(\omega)-S^{(2)}_{RL}(\omega)\right].~~
\eeq
The cross correlations 
\beq
S^{(2)}_{LR/RL}(\omega):=\int_{-\infty}^{\infty} d\tau e^{-i\omega\tau} \ew{\left\{ I_{L/R}(\tau)I_{R/L}(0) \right\} }_c,
\eeq
can be easily calculated taking the derivative of the CGF with respect to counting fields $\chi_L$ and $\chi_R$, while the particle-noise contributions $S^{(2)}_{L/R}$ involve a double derivative with respect to $\chi_L/\chi_R$ of the CGF.
Fig.~\ref{Fig1paperNMsup}c shows the noise associated with the charge fluctuations in the central island of an SET, $S^{(2)}_Q(\omega)$. Interestingly, if the SET is used as a detector of another quantum system, this noise governs the measurement backaction \cite{SchoelkopfCH, backaction1}. When $\hbar\omega\geq E_C$, the charge-noise spectrum contains extra quantum noise contributing to backaction, in full agreement with previous calculations \cite{Johansson,SchoelkopfCH}.

In Fig.~\ref{Fig2paperNMsup} we investigate this zero-frequency limit given by our NM theory. Fig.~\ref{Fig2paperNMsup}a shows the particle noise $S_R^{(2)}(\omega=0)$ as a function of voltage for a configuration such that $\varepsilon/\Gamma_R=80\gg T/\Gamma_R=4$. We observe a resonant step in the noise spectrum at precisely $V=\pm 2\varepsilon$. Above this step, there is a discrepancy of the Markovian solution with the NM and exact results, while right below the step, Markovian and non-Markovian limits differ from the exact solution. This last discrepancy is due to cotunneling contributions, only captured by the exact result. The difference is better observed in Fig.~\ref{Fig2paperNMsup}b, where we set $\varepsilon=0$ and vary the bias voltage again. Remarkably, the Markovian solution is flat for all voltages, while both NM and exact solutions show certain structure capturing system-bath memory effects. Only for low voltages these two disagree, when cotunneling contributions become important. At zero voltage, the Markovian and NM curves coincide as expected (since the only contribution to noise should originate from equilibrium fluctuations). For large enough voltages, the exact and NM results fall on top of each other, and we remark that the limit $V\to\infty$ is exact in both Markovian and non-Markovian approaches, and thus all three curves converge to the same value in this limit.

\subsection{Single-time full counting statistics}

\begin{figure}[t]
\centerline{
\includegraphics[width=3.3in]{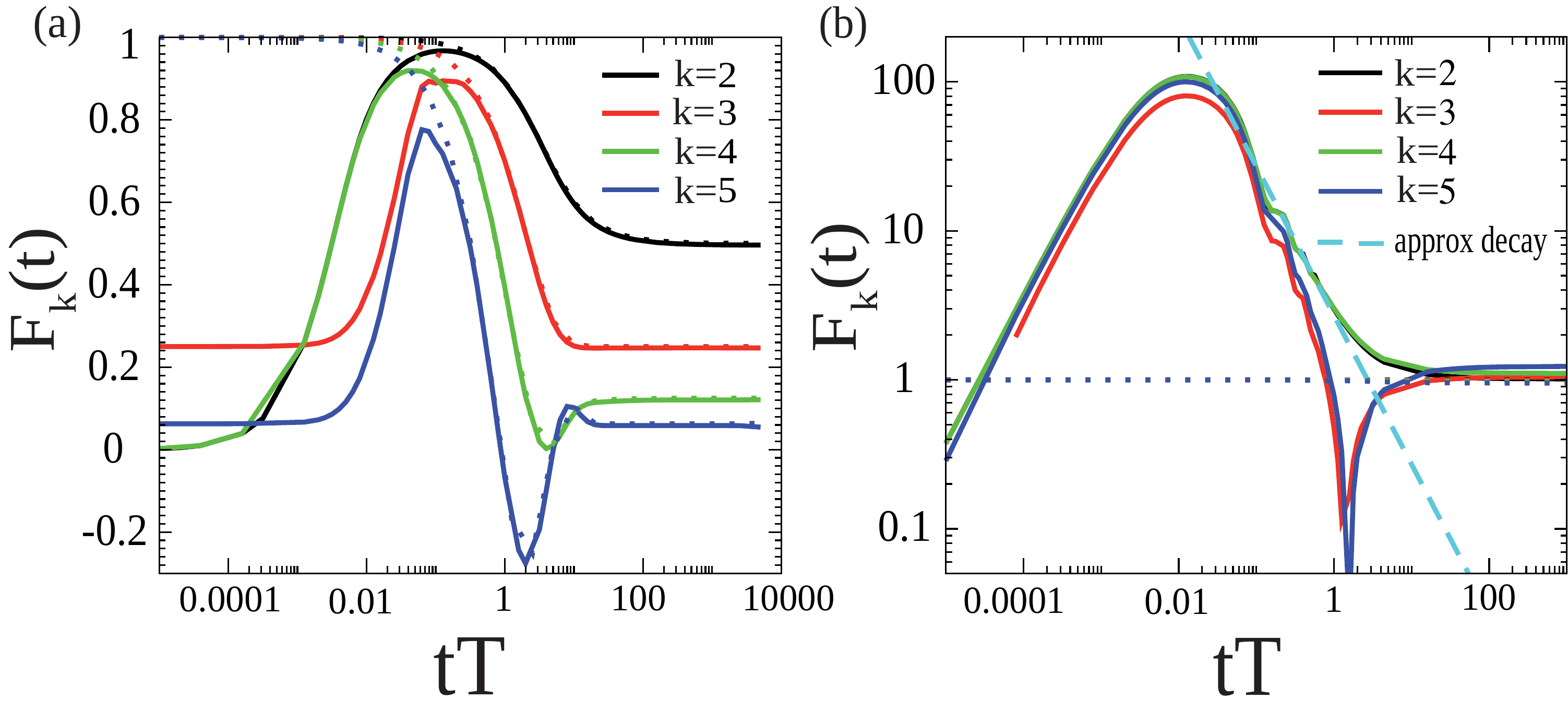}}
\caption{\label{Fig5}(color online). Single-time full counting statistics of the SRL model (Fano-factors of the right particle-current $F^{(k)}(t)\equiv\ew{n_R^k(t)}_c/\langle n_R\rangle$ with $\Gamma_L=\Gamma_R=0.25$, $T=1$) up to the fifth cumulant. a)~Shot noise regime ($\varepsilon=20$, $V=100$).  b)~Quantum fluctuations regime ($\varepsilon=20$, $V=30$). Dotted lines correspond to the Markovian approximation.}
\end{figure}

Beyond frequency-dependent noise spectra, Eq.~(\ref{inhomo2}) also allows us to study single-time full counting statistics of the number of electrons $n$ transferred to a particular terminal. This quantity is defined through the cumulant generating function ${\cal F}=\mathrm{log}\;{\cal G}$ as 
\beq \label{approxFCSt}
\ew{n^k(t)}_c = \frac{\partial^k}{\partial(i\chi)^k} \left. {\cal F}(\chi;t)\right|_{\chi=0}.
\eeq
Such $k$th-order cumulants can be measured by e. g. counting electrons using a quantum point contact and analyzing the time-dependent statistics of the events \cite{fli09}.  Fig.~\ref{Fig5}a shows the single-time Fano-factor $F^{(k)}\equiv \ew{n_R^k(t)}_c/\ew{n_R}$ of the SRL model. This figure shows up to the fifth order ($k=5$) Fano-factor (solid lines) together with the results corresponding to the MA (dotted lines) in the shot noise regime (level within the bias voltage window). At large times, the agreement with the Markovian solution is good for all the Fano-factors. At short times, however, the MA converges to the Poissonian limit while the NM solutions clearly show a strong sub-Poissonian supression. More interesting is when the level is above the bias window and all noise comes from quantum fluctuations (Fig.~\ref{Fig5}b). In this case, and taking an infinite bandwidth, the second cumulant $c_2(t)\equiv \ew{n^2(t)}_c$ can be approximated as the inverse Laplace transform of 
\beq
c_2(z)=z^{-2}\Gamma_R
\mathrm{Im}\left\{
  \frac{i}{2} + 
  \Psi\left(\frac{1}{2}-i\frac{(\varepsilon-\mu_R)+iz)}{2\pi k_BT}\right)
\right\},~~
\eeq 
with $\Psi$ the digamma function. This gives the exponentially large Fano-factor $F_2(t)\approx\frac{\Gamma_R}{2I_{\mathrm{dc}}}$ for very short times, and follows the power law $F_2(t)\approx\frac{\Gamma_R}{\pi(\varepsilon-\mu_R)I_{\mathrm{dc}}}t^{-1}$ at intermediate times. From this result we can estimate the time at which $F_2(t)$ deviates from the MA, namely $t_{\mathrm{switch}}=\frac{\Gamma_R}{\pi(\varepsilon-\mu_R)I_{\mathrm{dc}}}$. In Fig.~\ref{Fig5}b we plot this power law behavior (dashed blue line) together with the full NM solution (solid lines), and the Markovian solution, which here lie at the Poissonian value $1$. For times $t\ll t_{\mathrm{switch}}$, we obtain large super-Poissonian noise resulting from high-frequency quantum fluctuations.

\subsection{Double quantum dot}

To further illustrate the theory, we now consider the example of a double quantum dot (DQD). To the best of our knowledge, a complete study of this model in the different regimes of $V$, $T$ and $\omega$, and in the NM limit is yet lacking. The following results are also applicable to a Cooper pair box qubit. Again, the Markovian and NM solutions shown here correspond to first order in perturbation theory (sequential tunneling) and $S^{(2)}(\omega)$ refers to the `total' noise. In the Coulomb blockade regime, the possible DQD states are $|0\rangle\equiv|N_L,N_R\rangle$, $|L\rangle=|N_L+1,N_R\rangle$ and $|R\rangle=|N_L,N_R+1\rangle$, with $N_L$/$N_R$ being the number of electrons in the left/right dot. The qubit, with Hamiltonian ${\cal H}_\mathrm{S}=\varepsilon\left( \ket{L}\bra{L}-\ket{R}\bra{R} \right)+T_c \left( \ket{L}\bra{R}+\ket{R}\bra{L} \right)$, has eigenvalues $E_{\pm}=\pm{\Delta\over 2}$, being $\Delta\equiv2\sqrt{\varepsilon^2+T_c^2}$.
Near linear response ($eV\ll k_B T, \hbar\omega$), the only noise contribution originates from equilibrium fluctuations -- either thermal noise for $k_BT\gg\hbar\omega$, or quantum noise for $\hbar\omega\gg k_BT$. 
\begin{figure}[t]
\center
\includegraphics[width=2.75in]{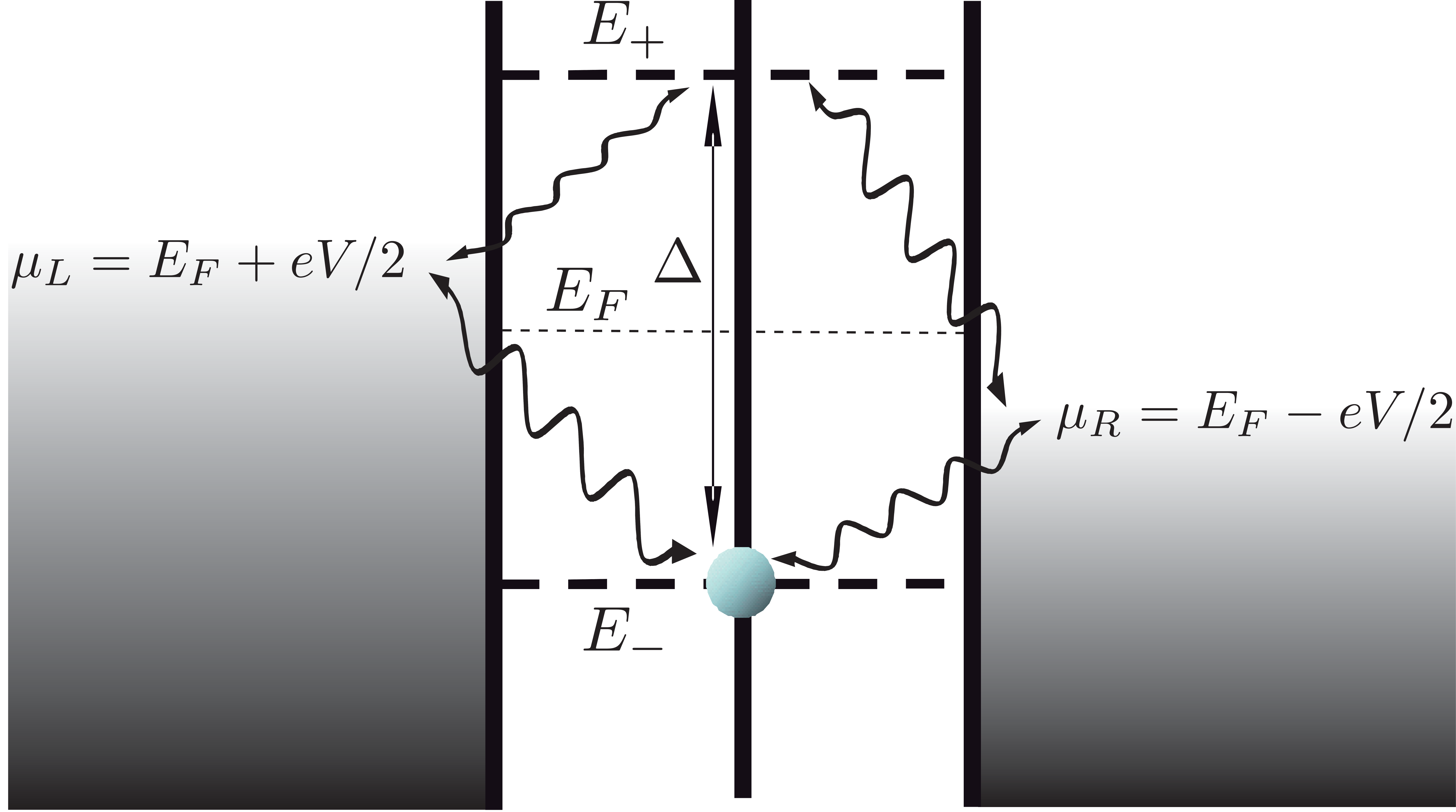}
\caption[Quantum noise processes in a double quantum dot]{(color online). Quantum noise processes in a double quantum dot. In the QNR, quantum fluctuations 
can discharge the system through the left/right reservoir if $\hbar\omega\geq |\mu_{L/R}-\Delta/2|$. These correspond to the steps in Fig.~\ref{Fig3paperNM}a. When $\omega=\Delta$, quantum interference between the eigenstates gives a noise suppression.}
\label{Fig2paperNM}
\end{figure}
In Fig.~\ref{Fig2paperNM} we sketch the physical processes due to quantum fluctuations, which give rise to the noise spectrum in Fig.~\ref{Fig3paperNM}a. 
For $eV\lesssim\Delta$, the conductance is zero and therefore $S^{(2)}(0)=0$, as dictated by the fluctuation-dissipation theorem. 
Quantum fluctuations, on the other hand, give rise to a finite noise for $\omega> 0$ (steps at $\hbar\omega=|\mu_{L/R}\pm\frac{\Delta}{2}|$ in Fig.~\ref{Fig3paperNM}a). 
Importantly, this physics is not captured with the MA, neither by other models for the inhomogeneity, such as $\Gamma(\chi; z)= \frac{1}{z}\{\Sigma(\chi,0)-\Sigma(\chi,z)\}$.
The spectrum also contains a strong dip centered at $\omega=\Delta$. This dip, which is voltage-independent and reaches $S^{(2)}(\omega=\Delta)=0$, can be understood as resulting from coherent destructive interference between the qubit eigenstates. This is demonstrated in Fig.~\ref{Fig3paperNM}b, where we investigate how this feature at $\omega=\Delta$ changes as we move the Fermi energy, $E_F$, of the reservoirs. For $V=0.1$ and $E_F=0$ (black solid curve), $E_{+/-}$ is above/below the chemical potentials and we find a dip shape, as discussed. When $E_F$ is aligned with the lowest level, namely $E_F=E_-=-\frac{\Delta}{2}$, the resonance changes to a Fano shape, as one expects from interference between a discrete level (the one above the chemical potentials at $E_+=\frac{\Delta}{2}$) and one strongly coupled to a continuum (the one at $E_F=E_-=-\frac{\Delta}{2}$). When both levels are above $E_F$, the interference at $\omega=\Delta$ is suppressed (red dotted curve). However, if both levels lie above $E_F$ (light grey curve), quantum interference still occurs, giving in this case a narrow resonant peak in the noise spectrum, since now we have a qubit weakly coupled to the leads -- therefore with a low dephasing rate. A very important remark of this figure, is that the situation corresponding to $E_F=-4$ gives a different result from that corresponding to $E_F=4$. In the former, the peak at $\omega=\Delta$ has been suppressed, while in the last, the resonance occurs. This we understand in terms of coherent oscillations only taking place when the levels lie below the chemical potentials. Most importantly, the light-grey curve only presents one quantum noise step, corresponding to the anti-bonding state. As the charge oscillates fast between both eigenstates, this can decay to the reservoirs via quantum noise processes only from the lowest level. However, in the situation with both eigenstates above the chemical potentials, charge can decay to the reservoirs from both levels through quantum noise processes. 
\begin{figure*}[t]
\center
\includegraphics[width=\textwidth] {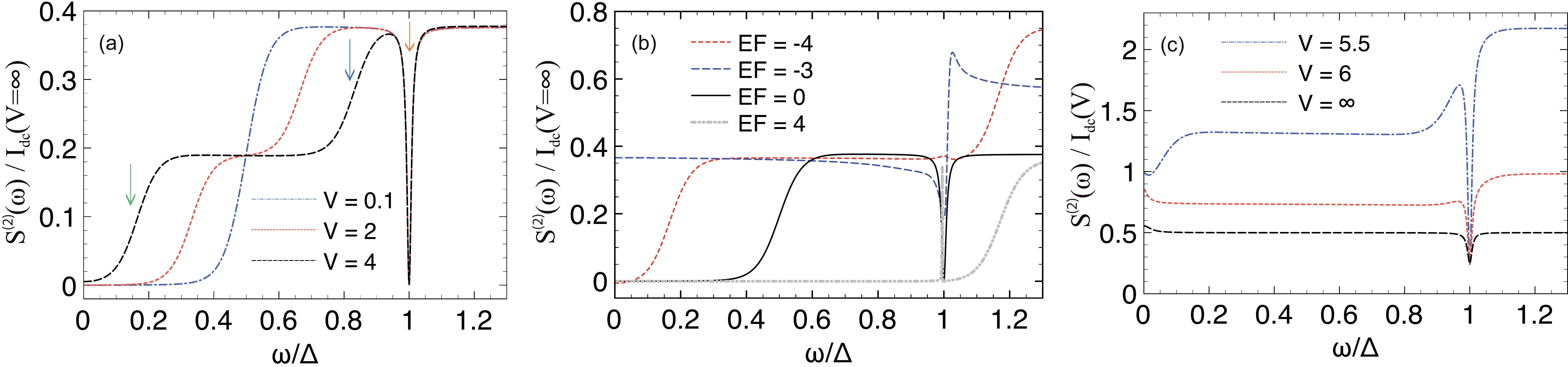}
\caption[Quantum noise spectra of a double quantum dot]{(color online). Finite-frequency noise of a double quantum dot (results normalized to the dc current  in the large-voltage limit $I_{\mathrm{dc}}(V=\infty)=t_c^2\Gamma_R/[\Gamma_R^2/4+4\varepsilon^2+t_c^2(2+\Gamma_R/\Gamma_L)]$. a)~Near linear response, $S^{(2)}(\omega)$ shows quantum noise steps at $\omega=|\Delta/2\pm V/2|$ and a dip centered at $\omega=\Delta$ (indicated with arrows for the case $V=4$ in the figure). b)~The feature at $\omega=\Delta$ originates from quantum interference between the bonding and anti-bonding qubit states. Here we set $V=0.1$ and vary the reservoir Fermi energy $E_F$, observing a displacement of the quantum noise step, as well as a modification of the resonance form at the qubit frequency $\Delta$ (see text). c)~Shot noise limit. Quantum noise steps are only visible for $V\lesssim \Delta, k_BT$, otherwise the contribution from shot noise or thermal noise are dominant. Parameters: $\varepsilon=0$, $\Delta=2T_c=6$, 
$\Gamma_L=\Gamma_R=T/2=0.1$.
}
\label{Fig3paperNM}
\end{figure*}
\begin{figure}[b]
\center
\includegraphics[width=2.2in] {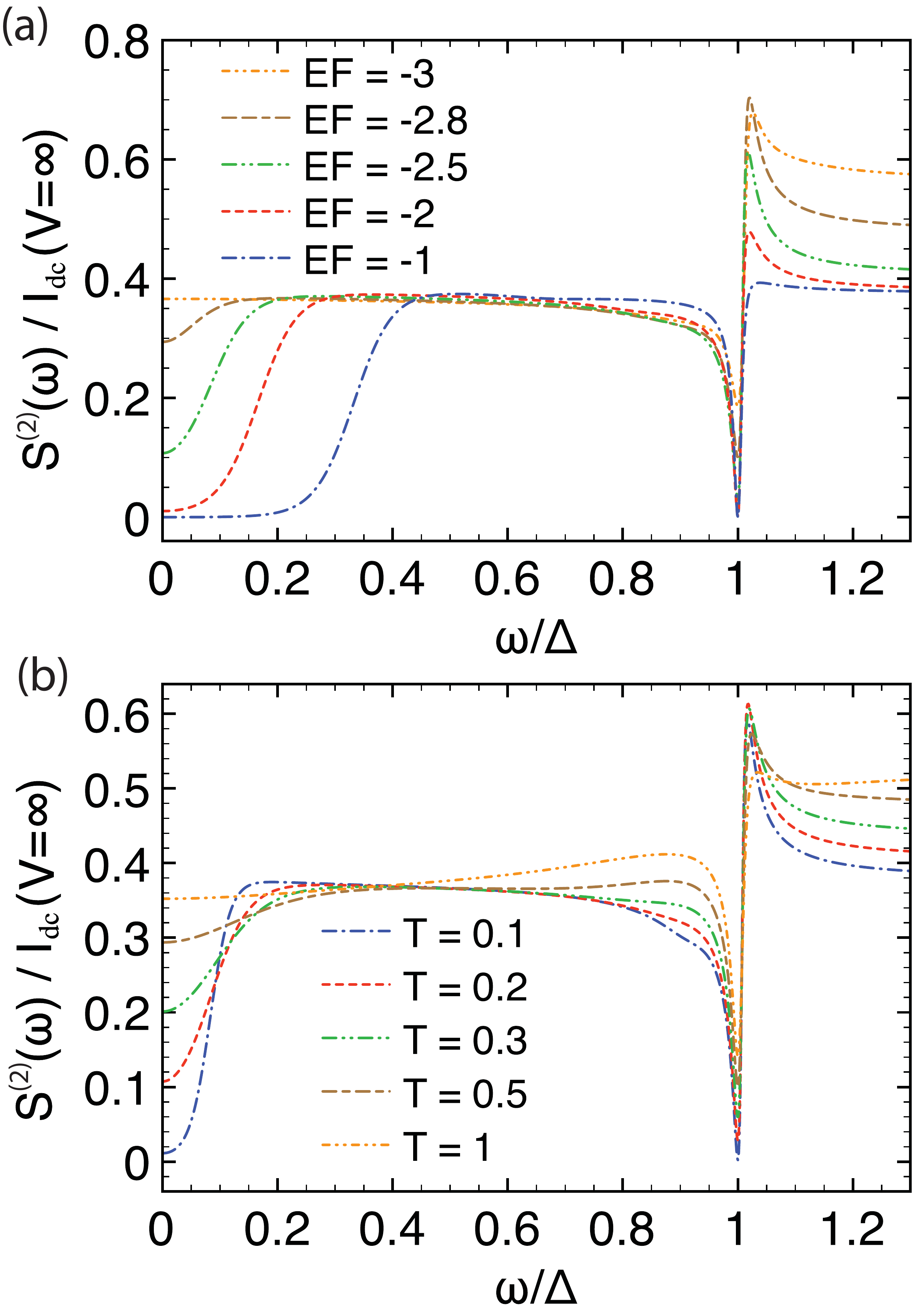}
\caption[Gate voltage and temperature effects on the quantum noise of a double quantum dot]{(color online). a)~Effect of a gate voltage. As the relative distance between the dot levels and the lead chemical potentials is varied (here illustrated decreasing the Fermi energy $E_F$), a quantum noise step, absent when the bonding state is aligned with both chemical potentials, appears at the corresponding frequency difference. The Fano shape, however, gives an anti-resonance at the qubit frequency $\Delta$. Here $T=0.2$. b)~Effect of the temperature. As $T$ is increased, the quantum noise step is lost, since thermal noise overcomes quantum noise, giving a finite $S^{(2)}$ value at zero frequency. The Fano shape is however preserved for high temperatures. Here $E_F=-2.5$. In both figures $V=0.1$, $\varepsilon=0$, $\Delta=2T_c=6$, $\Gamma_L=\Gamma_R=0.1$.}
\label{Fig4paperNMad}
\end{figure}
If $eV\gtrsim\Delta$, transport is possible and shot noise is finite, therefore $S^{(2)}(0)\neq 0$. This limit is discussed in Fig.~\ref{Fig3paperNM}c. Interestingly, quantum noise is progressively overcome by shot noise as $V$ increases. As a result, for large voltages, the quantum noise steps disappear and the noise is of smaller magnitude.
In this case an incomplete destructive interference is found at $\omega=\Delta$: $S^{(2)}(\omega=\Delta)/I_{\mathrm{dc}}(V)$ is greater than zero and does not depend on $V$. 
The width, on the other hand, increases with the voltage, which can be understood as a decrease of the dephasing time (inverse of the width) due to the coupling with the reservoirs \cite{Aguado-Brandes}. The MA is recovered as $V\rightarrow\infty$, with features at $\omega=0$ and $\omega=\Delta$ on top of a background of sub-Possonian partition noise, Fano-factor $S^{(2)}(\omega)/I_\mathrm{dc}(V)=1/2$.

The transition from a Fano shape to an anti-resonance in the noise spectrum encountered in Fig.~\ref{Fig3paperNM}b is further investigated in figure \ref{Fig4paperNMad}a. Here we show how the quantum noise step progressively appears as the bonding state comes below the chemical potentials. At the same time, the Fano resonance gives rise to the destructive-interference feature at the qubit frequency. The effect of temperature is shown in Fig.~\ref{Fig4paperNMad}b. Still in the linear response regime, where the `shot' contribution is negligible, we see how quantum noise is overcome by thermal noise, giving a finite $S^{(2)}$ value at zero frequency for increasing temperature, as dictated by the fluctuation-dissipation theorem. The Fano shape, consequence of having the lowest level strongly coupled to the reservoirs, but also coupled to the anti-bonding state, persists at high temperatures.

\section{Conclusions}

We have presented a general non-Markovian theory of frequency-dependent noise based on QME. The importance of NM correlations to correctly capture the physics of vacuum fluctuations has been shown through the study of a single resonant level model and a double quantum dot in the quantum noise regime. Our equations for the CGF and noise spectrum open the possibility to investigate this physics in a variety of systems where NM corrections are of vital importance, such as electromechanical resonators close to the zero-point motion \cite{Connell10}, or strongly correlated cold atoms in optical lattices \cite{Braungardt10}.

We gratefully acknowledge C. Flindt and A. Braggio for discussions. Work supported by MICINN-Spain (Grants FIS2009-08744, FPU AP2005-0720), Acci\'on Integrada Spain-Germany Grant HA2007-0086, the WE Heraeus foundation, and by DFG (grant BR 1528/5-2).

\appendix

\section{Liouvillian Perturbation Theory} \label{appKernel}

For concrete application of the formalism, we employ Liouvillian perturbation theory (LPT), as described in Refs.~[\onlinecite{Schoeller09,CE09cot,CE10SCLPT1}].
We give here a brief review of the essential elements of this theory --- for more details, the reader is referred to the original references. As explained in the main text, the Hamiltonian describing single-electron tunnelling between central system and reservoirs is
\beq
  {\cal H}_T =  \sum_{k \alpha m} 
  V_{k \alpha m} c^\dag_{k\alpha} d_m 
  + V^*_{k \alpha m} d^\dag_m c_{k\alpha}
  \label{V1}
  ,
\eeq
where $d_m$ is the annihilation operator for the single-particle level $m$ in the central system, and $c_{k\alpha}$ the annihilation operator for an electron with momentum $k$ in lead $\alpha$.
and $V_{k \alpha m}$ is a tunnelling amplitude. 
Introducing a compact single index ``$1$'' to denote the indices $(\xi_1,k_1,\alpha_1)$, we have
\beq
  c_1 =  c_{\xi_1 k_1 \alpha_1}
  =
  \left\{ 
    \begin{array}{c c}
      c^\dag_{k_1\alpha_1}, &\quad \xi_1 =+\\
      c_{k_1\alpha_1}, &\quad \xi_1 =-
    \end{array}
  \right. 
  ,
\eeq
with index $\xi_1=\pm$ indicating whether the operator is a creation or annihilation operator.
We further define the system operators
$
  g_{k\alpha} \equiv
  \sum_{m}  V_{k \alpha m} d_{m}
  \label{gtj}
$,
such that the tunnel Hamiltonian can be simply written as
$
  {\cal H}_T = c_1 g_1
$,
where $\overline{1}$ denotes $(-\xi_1,k_1,\alpha_1)$, and here as elsewhere, implicit sums over repeated indices. In the same notation, the reservoir Hamiltonian ${\cal H}_{\mathrm{res}}=\sum_{k,\alpha\in L,R} \varepsilon_{k\alpha} c_{k\alpha}^{\dagger}c_{k\alpha}$ reads 
$
  {\cal H}_\mathrm{res} =
  \varepsilon_1 c_1 c_{\overline{1}}\delta_{\xi_1+}
$,

In Liouville space the tunneling Liouvillian can be written as
\beq
  {\cal L}_T = -i\left[H_T,\bullet~\right]
  =
  -i \xi_1 \sum_p p \sigma^p C^p_1 G^p_1
  ,
\eeq
where $p=\pm$ is a Keldysh index corresponding to the two parts of the commutator. $C$ and $G$ are the superoperators corresponding to $c$ and $g$, defined through their actions on arbitrary operator $O$:
\beq
  C^p_1 O &=& 
  \left\{
   \begin{array}{c c}
      a_1 O, &p =+ \\
     O a_1, &p =-
    \end{array}
  \right.
  \label{ddef}
  ,\\
  G^p_1 O &=& 
  \sigma^p 
  \times
  \left\{
   \begin{array}{c c}
      g_1 O, &p =+ \\
      - O g_1, &p =-
    \end{array}
  \right.
  \label{Gdef}
  .
\eeq
The object $\sigma^p$ is a dot-space superoperator with matrix elements
$
  \rb{\sigma^p}_{ss',\bar{s}\bar{s}'} = 
  \delta_{s\bar{s}}\delta_{s'\bar{s}'}  p^{N_s-N_{s'}}
$ where $N_s$ is the number of electrons in state $s$.

We can now express the self-energy in terms of these superoperators. As described in the main text, the non-Markovian system self-energy can be obtained by expansion of 
\beq
  {\rho}_\mathrm{S}(z)
  =\mathrm{Tr}_\mathrm{res}
  \left\{
    \left[z-{\cal L}\right]^{-1}
    \rho_\mathrm{S}(t_0) \otimes \rho_\mathrm{res}^\mathrm{eq}
  \right\}
  =\Omega(z){\rho}_\mathrm{S}(t_0)\nonumber,
\eeq 
as a power series of ${\cal L}_T$ and tracing out bath degrees of freedom. This can be done using the diagrammatic technique explained in Refs.~[\onlinecite{Schoeller09,CE09cot,CE10SCLPT1}]. To lowest order (sequential), we obtain
\beq
  \Sigma(z) &=& -
  G^{p_2}_2 
  \Omega_\mathrm{S}(z -i\xi_2\varepsilon_2)
  G^{p_1}_1 \gamma_{21}^{p_2 p_1}
  \label{sig2translate}
  .
\eeq
In this expression we find the free-propagator
$
   \Omega_\mathrm{S}(z) 
   = \frac{1}{z-{\cal L}_\mathrm{S}}
$, and the reservoir contraction
\beq
  \gamma_{21}^{p_2p_1}=\delta_{2\overline{1}} p_1 f(-\xi_1 p_1 \varepsilon_1),
\eeq
with Fermi function $f(\varepsilon_{\alpha})=(e^{\varepsilon-\mu_{\alpha}}+1)^{-1}$.

Counting in lead $\alpha$ is introduced through the replacement $V_{k \alpha m}\to V_{k \alpha m} e^{i p \chi_{\alpha}/2}$ in the tunnel Liouvillian ${\cal L}_\mathrm{T}$.
The $\chi$-dependent self-energy is then simply obtained as the above self-energy but with $\chi$-dependent superoperator $G(\chi)$ replacing $G$. 
The two-point self-energy determining $\Gamma(\chi,z)$ can similarly be derived. We obtain
\beq
  \Pi(\chi_2,\chi_1,z)
  &=&
   - 2\pi p_1 G^{p_2}_{\bar{1}}  \opp{\phi_a}{\phi_a} G^{p_1}_1
   \nonumber\\
   &&
   \times
   e^{
     i \frac{1}{2} \delta_{\alpha_1\beta}  \xi_1 
     (p_1 \chi_1-p_2 \chi_2)
   }
   \nonumber\\
   &&
   \times
   I_{p_1}^{(2)}(\Delta_a + \xi_1 \mu_{\alpha_1} -i z)
   ,
\eeq
with $\Delta_a$, $\kett{\phi_a}$, eigenvalues and eigenvectors of the central-system Liouvillian, that is ${\cal L}_\mathrm{S} \kett{\phi_a} = -i \Delta_{a}\kett{\phi_a} $, and $I^{(2)}_p(\lambda)$ defined as
\beq
  I_{p}^{(2)}(\lambda) 
   \equiv\frac{i}{2\pi} 
	\int d \omega
	\frac{f(\omega)}{i0^+ + p \omega - \lambda},
\eeq
The upper limit of this integral can be taken as a Lorentzian cutoff $D(\omega)=X_c^2/(\omega^2 + X_c^2)$, which gives
\beq 
  I_{p}^{(2)}(\lambda) &=& D(\lambda)
  \left\{
    \frac{1}{2}f(p\lambda) + \frac{i p }{2 \pi} \phi(\lambda) - \frac{i \lambda}{4 X_c}
  \right\}.
  \label{Iseq2b} 
\eeq
Here, 
$
   \phi(\lambda) \equiv
   \frac{1}{2}
  \rb{\frac{}{}
  g(\lambda)+ g(-\lambda)
  -2 g(i X_c)
  }
$ and $g(\omega) \equiv \Psi\rb{\frac{1}{2} + \frac{\omega}{2\pi i}}$, being $\Psi(x)$ the digamma function.
In the wide-band limit ($X_c\gg\omega,\lambda$), this integral becomes
\beq
  I_{p}^{(2)}(\lambda) 
  &=&  \frac{1}{2}f(p\lambda) + \frac{i p }{2 \pi} \phi(\lambda) 
  \label{IlargeXC2}
   ,
\eeq
with approximate $\phi$-function 
\beq
   \phi(\lambda) = \frac{1}{2}\rb{g(\lambda)-g(-\lambda)}-\log(X_c/2\pi)
   \label{phiapp}
   .
\eeq
This latter result is adequate for finite-frequency shotnoise calculations, but the more accurate form \eq{Iseq2b} is required to correctly capture the the single-time full counting statistics, for which a bandwidth $X_c=500 k_BT$ was assumed.


\begin{thebibliography}{99}
\bibitem{steady} Here $\langle \hat{I}(t+\tau)\hat{I}(t) \rangle_c\equiv\langle \hat{I}(t+\tau)\hat{I}(t) \rangle- \langle \hat{I}(t+\tau)\rangle\langle\hat{I}(t) \rangle$ is the usual cumulant average and $\{\ldots\}$ is the anticommutator. Notice that for systems with no explicit time-depencence, or driving, $\ew{I(t)}$ can be replaced by $I_{\mathrm{dc}} = \lim_{t\to\infty} \ew{I(t)}$.
\bibitem{alphabeta} Here $\alpha$ and $\beta$ describe how the current is partitioned. According to the Ramo-Shockley theorem, for a two-terminal conductor the total current can be written in terms of the currents through the left and right terminals as: $I_{\mathrm{tot}}=\alpha I_L + \beta I_R$.
\end{thebibliography}
\end{document}